\documentstyle[aps,prd,multicol]{revtex}
\tighten

\def\Lrule{\vspace*{-0.2in}\noindent\vrule width3.4in height.2pt
  depth.2pt \vrule depth0em height.5em}
\def\Rrule{\vspace{-0.1in}\hfill\vrule depth.5em height0pt \vrule
  width3.4in height.2pt depth.2pt\vspace*{-0.1in}}

\begin{document}

\title{Compactification of gauge theories and the gauge
invariance of massive modes}

\author{R. Amorim$^a$ and J. Barcelos-Neto$^b$}

\address{\mbox{}\\
Instituto de F\'{\i}sica\\
Universidade Federal do Rio de Janeiro\\
RJ 21945-970 - Caixa Postal 68528 - Brasil}
\date{\today}

\maketitle
\begin{abstract}
\hfill{\small\bf Abstract\hspace*{1.7em}}\hfill\smallskip
\par
\noindent
We study the gauge invariance of the massive modes in the
compactification of gauge theories from $D=5$ to $D=4$. We deal with
gauge theories of rank one and two. In the compactification of the
Maxwell theory, St\"uckelberg fields naturally appear in order to
render gauge invariance for the massive modes. We show that in the
rank two case, where compactification leads to massive and massless
rank two and rank one modes, respectively, the vector modes play the
role of St\"uckelberg fields for the rank two modes. We also show that
in the non-Abelian case (where just rank one is considered)
St\"uckelberg fields appear in a very different way when compared with
their usual implementation in the non-Abelian Proca model.
\end{abstract}

\smallskip\mbox{}

\begin{multicols}{2}
\section{Introduction}
\renewcommand{\theequation}{1.\arabic{equation}}

Nowadays, there is broad consensus that fundamental theories might
come from spacetime dimension $D$ higher than four, probably $D=10$ or
$D= 11$. This is mainly related to the advent of string theories, that
are consistent in the quantum word just at $D=10$. Further, the
duality among known string theories suggests that they might emerge
from a more fundamental theory at $D=11$. However, one of the great
drawbacks of this idea is that there is no rule about some specific
mechanisms to reach our word at $D=4$.

\medskip
One manner of gaining some insight on this problem is to study the
compactification close to $D=4$ and try to understand the features
that a possible fundamental theory should have from the theoretical
consistency of the results. It is important to mention that the
seminal idea related to this point of view dates back a long time ago,
in the works of Kaluza and Klein \cite{Kaluza,Castellani}, where they
have started from the gravitational Einstein theory at $D= 5$ and,
after spontaneous compactification, reached the Maxwell and Einstein
theories at $D=4$. The vector gauge field $A^\mu$ was originated from
a component of the metric tensor. It is opportune to emphasize that
many of the recent attempts to implement compactification mechanism
have the Kaluza-Klein idea as a strong support.

\medskip
We would like to address the present paper to this line. We consider
the compactification of gauge theories of ranks one and two  from $D=
5$ to $D=4$ and study the question of gauge invariance in $D=4$, that
shall be retained in the compactification procedure, even for the
massive modes. The St\"uckelberg fields \cite{St}, that are usually
introduced as a trick in order to attain the gauge invariance of
massive vector fields, naturally emerge in this procedure. Even though
we are going to deal just with gauge fields of rank one and two (that
are the only ones whose number of physical degrees of freedom are
consistent at $D=4$), the method can be directly extended for gauge
fields of any rank at higher spacetime dimensions. Concerning to the
non-Abelian case, we mention that the non-Abelian formulation of
theories of rank higher than one cannot be directly done. The gauge
invariance, even for the massless case, is only achieved by means of
auxiliary fields \cite{Lahiri}. We shall deal with the non-Abelian
case just for rank one. We mention that the role played by the
St\"uckelberg fields in this case is very different from their usual
implementation in the non-Abelian Proca model.

\medskip
Our work is organized as follows: In Section II we review the
spontaneous compactification of Maxwell theory originally defined in
$D=5$. In the Fourier expansion procedure, the zero modes correspond
to Maxwell and real massless scalar fields in $D= 4$. The nonzero
modes correspond to complex Proca fields coupled to the appropriate
complex St\"uckelberg fields to keep the gauge invariance
\cite{Castellani,Dienes}. In Section III we apply this procedure in
the two-form Abelian gauge theory originally described in $D=5$. After
compactification, we get Maxwell and massless two-form gauge theories
for the zero modes. The vector fields play the role of St\"uckelberg
fields for the massive modes of the two-form fields. In section IV, we
consider Yang-Mills theory. Although it has the Abelian limit found in
Section II, an interesting gauge structure is obtained in the full
theory, where modes and gauge multiplets are mixed in a non-trivial
way in order to keep the gauge invariance of the action. We reserve
Section V for some concluding remarks.

\vfill\newpage

\section{Maxwell theory}
\renewcommand{\theequation}{2.\arabic{equation}}
\setcounter{equation}{0}

Let us briefly review the compactification of the Maxwell theory. The
corresponding action is

\begin{equation}
S=\frac{1}{R}\int d^4x\int_0^Rdx^4\,
\Bigl(-\,\frac{1}{4}F^{MN}F_{MN}\Bigr)
\label{2.1}
\end{equation}

\noindent
where  the coordinate $x^4$ describes a circle of radius $R$. We use
capital Roman indices to express the spacetime dimension $D=5$, i.e.
$M,N=0,\cdots,4$, and adopt the metric convention $\eta^{MN}=
diag\,(+1,-1,-1,-1,-1)$. The Maxwell stress tensor $F^{MN}$ is defined
in terms of the potential vector $A^M$ by the usual relation

\begin{equation}
F^{MN}=\partial^MA^N-\partial^NA^M
\label{2.2}
\end{equation}

\noindent
The action (\ref{2.1}) is invariant under the
gauge transformation

\begin{equation}
\delta A^M=\partial^M\Lambda
\label{2.3}
\end{equation}

Let us split the vector potential $A^M$ as $A^M=A^\mu$ for $M=
0,\cdots,3$ and $A^M=\phi$ for $M=4$. We thus have for the action
(\ref{2.1})

\begin{eqnarray}
&&S=\frac{1}{R}\int d^4x\int_0^Rdx^4
\Bigl(-\,\frac{1}{4}F^{\mu\nu}F_{\mu\nu}
-\frac{1}{2}\partial^\mu\phi\,\partial_\mu\phi
\nonumber\\
&&\phantom{S=\frac{1}{R}\int d^4x\int_0^R}
-\frac{1}{2}\partial^4 A^\mu\partial_4A_\mu
+\partial^\mu\phi\,\partial^4A_\mu\Bigr)
\label{2.4}
\end{eqnarray}

\noindent
The first term above cannot be identified with the Maxwell Lagrangian
at $D=4$ because $A^\mu$ depends on both $x^\mu$ and $x^4$. Following
the usual procedure in the (spontaneous) compactification procedure
\cite{Castellani,Dienes}, we take the expansions of $A^\mu$ and
$\phi$ in Fourier harmonics, namely

\begin{eqnarray}
A^\mu(x,x^4)&=&\sum_{n=-\infty}^{+\infty}A_{(n)}^\mu(x)\,
\exp\Bigl(2in\pi\frac{x^4}{R}\Bigr)
\nonumber\\
\phi(x,x^4)&=&\sum_{n=-\infty}^{+\infty}\phi_{(n)}(x)\,
\exp\Bigl(2in\pi\frac{x^4}{R}\Bigr)
\label{2.5}
\end{eqnarray}

\noindent
The replacement of these expansions into the action (\ref{2.4}) leads
to

\begin{eqnarray}
&&S=\int d^4x\sum_{n=-\infty}^{+\infty}
\Bigl(-\,\frac{1}{4}F_{(n)}^{\mu\nu}F_{(-n)\mu\nu}
-\frac{1}{2}\partial^\mu\phi_{(n)}\partial_\mu\phi_{(-n)}
\nonumber\\
&&\phantom{S=\int d^4x}
-\frac{2\pi^2n^2}{R^2}\,A_{(n)}^\mu A_{(-n)\mu}
+\frac{2in\pi}{R}\,A_{(n)}^\mu\partial_\mu\phi_{(-n)}\Bigr)
\nonumber\\
\label{2.6}
\end{eqnarray}

Since $A_M$ is a real quantity, we have from expansions (\ref{2.5})
that $A_{(-n)}^\mu=A_{(n)}^{\ast\mu}$ and $\phi_{(-n)}=\phi_{(n)}
^\ast$. Using this into the action (\ref{2.6}) we rewrite it in a more
convenient way

\end{multicols}
\renewcommand{\theequation}{2.\arabic{equation}}
\Lrule

\begin{eqnarray}
&&S=\int d^4x\Bigl\{-\,\frac{1}{4}\,F_{(0)}^{\mu\nu}F_{(0)\mu\nu}
-\frac{1}{2}\,\partial^\mu\phi_{(0)}\partial_\mu\phi_{(0)}
\nonumber\\
&&\phantom{S=\int d^4x\Bigl\{}
+\sum_{n=1}^\infty\Bigl[
-\,\frac{1}{2}F_{(n)}^{\mu\nu}F_{(n)\mu\nu}^\ast
-\,\frac{4n^2\pi^2}{R^2}
\Bigl(A_{(n)}^\mu+\frac{iR}{2n\pi}\,\partial^\mu\phi_{(n)}\Bigr)
\Bigl(A_{(n)\mu}^\ast-\frac{iR}{2n\pi}\,
\partial_\mu\phi_{(n)}^\ast\Bigr)\Bigr]\Bigr\}
\label{2.7}
\end{eqnarray}

\bigskip
\Rrule
\begin{multicols}{2}

We observe that the two zero mode terms at $D=4$ correspond to Maxwell
and real scalar field theories. The other modes are related to complex
Proca fields with masses given by $2n\pi/R$ and to massless complex
scalar fields. These are St\"uckelberg fields. Usually, they are put
by hand as a trick to make the Proca theory gauge invariant or to
implement the Hamiltonian embedding procedure, during the conversion
of second to first-class constraints \cite{BFFT}. Here, these fields
naturally emerge in order to keep the gauge symmetry of the initial
theory.

\section{Abelian two-form}
\renewcommand{\theequation}{3.\arabic{equation}}
\setcounter{equation}{0}

Let us extend what was done in the previous section by considering
Abelian gauge fields of rank two. The corresponding action reads
\cite{Kalb}

\begin{equation}
S=\frac{1}{12R}\int d^4x\int_0^Rdx^4\,H^{MNP}H_{MNP}
\label{3.1}
\end{equation}

\noindent
The completely antisymmetric stress tensor $H^{MNP}$ is defined in
terms of the antisymmetric two-form gauge field $B^{MN}$ by

\begin{equation}
H^{MNP}=\partial^MB^{NP}
+\partial^PB^{MN}+\partial^NB^{PM}
\label{3.2}
\end{equation}

\noindent
This theory is invariant for the (reducible) gauge transformation
\cite{Kalb,Henneaux}

\begin{equation}
\delta B^{MN}=\partial^M\Lambda^N-\partial^N\Lambda^M
\label{3.3}
\end{equation}

In order to perform the compactification to $D=4$, we conveniently
split the potential $B^{MN}$ as

\begin{eqnarray}
B^{MN}&=&(B^{\mu\nu},B^{4\mu})
\nonumber\\
&=&(B^{\mu\nu},A^\mu)
\label{3.4}
\end{eqnarray}

\noindent
where we have identified $B^{4\mu}$ with $A^\mu$. Again, this is not
the vector potential of the Maxwell theory because it depends on both
$x^\mu$ and $x^4$ and its gauge transformation, according to
(\ref{3.3}), reads

\begin{equation}
\delta A^\mu=\partial^\mu\Lambda^4-\partial^4\Lambda^\mu
\label{3.5}
\end{equation}

\noindent
which is not the characteristic transformation of the Maxwell
connection.

\medskip
Introducing (\ref{3.4}) into (\ref{3.1}), we obtain

\begin{eqnarray}
&&S=\frac{1}{R}\int d^4x\int_0^Rdx^4
\Bigl(\frac{1}{12}\,H^{\mu\nu\rho}H_{\mu\nu\rho}
-\frac{1}{4}\,F^{\mu\nu}F_{\mu\nu}
\nonumber\\
&&\phantom{S=\frac{1}{R}\int d^4x}
+\frac{1}{4}\,\partial^4B^{\mu\nu}\partial_4B_{\mu\nu}
+\frac{1}{2}\,F^{\mu\nu}\partial^4B_{\mu\nu}\Bigr)
\nonumber\\
\label{3.6}
\end{eqnarray}

\noindent
where, for the same previous argument, $F^{\mu\nu}=\partial^\mu
A^\nu-\partial^\nu A^\mu$ is not the Maxwell stress tensor. Expanding
the fields $B^{\mu\nu}$, $A^\mu$, as well as the gauge parameters
$\Lambda^\mu$ and $\Lambda^4$ in Fourier harmonics, we have

\begin{eqnarray}
B^{\mu\nu}(x,x_4)&=&\sum_{n=-\infty}^{+\infty}
B_{(n)}^{\mu\nu}(x)\,\exp\Bigl(2in\pi\frac{x^4}{R}\Bigr)
\nonumber\\
A^\mu(x,x_4)&=&\sum_{n=-\infty}^{+\infty}
A_{(n)}^{\mu}(x)\,\exp\Bigl(2in\pi\frac{x^4}{R}\Bigr)
\nonumber\\
\Lambda^\mu(x,x_4)&=&\sum_{n=-\infty}^{+\infty}
\xi_{(n)}^\mu(x)\,\exp\Bigl(2in\pi\frac{x^4}{R}\Bigr)
\nonumber\\
\Lambda^4(x,x_4)&=&\sum_{n=-\infty}^{+\infty}
\xi_{(n)}(x)\,\exp\Bigl(2in\pi\frac{x^4}{R}\Bigr)
\label{3.7}
\end{eqnarray}

\noindent
Introducing these quantities into (\ref{3.6}) and considering that
$B^{\mu\nu}_{(-n)}=B_{(n)}^{\ast\mu\nu}$, $A_{(-n)}^\mu=A_{(n)}
^{\ast\mu}$, etc., we obtain

\begin{eqnarray}
&&S=\int d^4x\Bigl[\frac{1}{12}\,
H_{(0)}^{\mu\nu\rho}H_{(0)\mu\nu\rho}^\ast
-\frac{1}{4}\,F_{(0)}^{\mu\nu}F_{(0)\mu\nu}^\ast
\nonumber\\
&&\phantom{S=\int d^4x\Bigl[}
+\sum_{n=1}^\infty\Bigl(
\frac{1}{6}\,H_{(n)}^{\mu\nu\rho}H_{(n)\mu\nu\rho}^\ast
-\frac{1}{2}\,F_{(n)}^{\mu\nu}F_{(n)\mu\nu}^\ast
\nonumber\\
&&\phantom{S=\int d^4x\Bigl[}
+\frac{2n^2\pi^2}{R^2}\,B_{(n)}^{\mu\nu}B_{(n)\mu\nu}^\ast
-\frac{2in\pi}{R}\,F_{(n)}^{\mu\nu}B_{(n)\mu\nu}^\ast\Bigr)\Bigr]
\nonumber\\
\label{3.8}
\end{eqnarray}

Due to the gauge transformations of the zero mode fields,

\begin{eqnarray}
&&\delta A_{(0)}^\mu=\partial^\mu\xi_{(0)}
\nonumber\\
&&\delta B_{(0)}^{\mu\nu}=\partial^\mu\xi_{(0)}^\nu
-\partial^\nu\xi_{(0)}^\mu
\label{3.9}
\end{eqnarray}

\noindent
we have that the two zero mode terms of (\ref{3.8}) are the tensor and
vector (Maxwell) theories at $D=4$ \cite{Barc1}. The remaining modes
correspond to massive complex tensor gauge fields $B_{(n)}^{\mu\nu}$
and massless vector ones $A_{(n)}^\mu$. Let us rewrite the $n$-mode
terms of expression (\ref{3.8}) in a more appropriate form

\end{multicols}
\renewcommand{\theequation}{3.\arabic{equation}}
\Lrule

\begin{equation}
S_{(n)}=\int d^4x
\Bigl[\frac{1}{6}\,H_{(n)}^{\mu\nu\rho}H_{(n)\mu\nu\rho}^\ast
-\frac{2n^2\pi^2}{R^2}
\Bigl(iB_{(n)}^{\mu\nu}-\frac{R}{2n\pi}\,F_{(n)}^{\mu\nu}\Bigr)
\Bigl(-\,iB_{(n)\mu\nu}^\ast-\frac{R}{2n\pi}\,
F_{(n)\mu\nu}^\ast\Bigr)\Bigr]
\label{3.10}
\end{equation}

\bigskip
\Rrule
\begin{multicols}{2}

We notice that in the rank 2 theory, the massless vector gauge field
$A_{(n)}^\mu$ naturally appears as a St\"uckelberg field for the
massive antisymmetric gauge field $iB_{(n)}^{\mu\nu}$. Their gauge
transformations are given by

\begin{eqnarray}
&&\delta A_{(n)}^\mu=\partial^\mu\xi_{(n)}
-\frac{2i\pi n}{R}\,\xi_{(n)}^\mu
\label{3.11a}\\
&&\delta B_{(n)}^{\mu\nu}=\partial^\mu\xi_{(n)}^\nu
-\partial^\nu\xi_{(n)}^\mu
\label{3.11b}
\end{eqnarray}

\noindent
It is interesting to observe the expression (\ref{3.11a}), that gives
the gauge transformation for $A_{(n)}^\mu$. The first term is the
usual gauge transformation for vector fields. The second one is the
term related to its new role as St\"uckelberg field for the massive
rank two modes.

\medskip
We mention that a similar result, where vector fields play the role of
St\"uckelberg fields for the massive rank two theory, was also found
in the case of Hamiltonian embedding mechanism \cite{Bizdadea}.

\medskip
We could generalize this analysis for gauge fields of any rank.
However, for $D=4$ it does not make sense to consider gauge fields of
rank higher than two, because the gauge and redutibility conditions
would lead to a negative number of physical degrees of freedom.

\section{Non-Abelian case}
\renewcommand{\theequation}{4.\arabic{equation}}
\setcounter{equation}{0}

As already said, the non-Abelian formulation of gauge theories with
rank two (or higher) cannot be directly implemented. Its gauge
invariance can only be achieved with the help of auxiliary fields
\cite{Lahiri}. Consequently, since these models do not have a clear
gauge invariance, their discussion here will be avoided. We shall only
consider in this section the vector case. The non-Abelian version of
the procedure described in section II should start from the action

\begin{equation}
S=\frac{1}{R}\int d^4x\int_0^Rdx^4\,
\mbox{tr}\,\Bigl(-\,\frac{1}{4}F^{MN}F_{MN}\Bigr)
\label{4.1}
\end{equation}

\noindent
where now

\begin{equation}
F^{MN}=\partial^MA^N-\partial^NA^M-i[A^M,A^N]
\label{4.2}
\end{equation}

\noindent
and the gauge potentials  take values in a $SU(N)$ algebra, whose
hermitian generators are assumed to satisfy a normalized trace
condition. The action (\ref{4.1}) is invariant under the gauge
transformation

\begin{equation}
\delta A^M=D^M\Lambda
\label{4.3}
\end{equation}

\noindent
once one defines the covariant derivative as

\begin{equation}
D^M\Lambda=\partial^M\Lambda-i[\Lambda,A^M]
\label{4.4}
\end{equation}

\noindent
Again writing $A^4=\phi$, we note that $F^{4\mu}=\partial^4A^\mu-
D^\mu\phi$ permits to rewrite action (\ref{4.1}) as

\begin{eqnarray}
&&S=\frac{1}{R}\int d^4x\int_0^Rdx^4\,
\mbox{tr}\,\Bigl(-\,\frac{1}{4}F^{\mu\nu}F_{\mu\nu}
\nonumber\\
&&\phantom{S=\frac{1}{R}\int}
-\frac{1}{2}D^\mu\phi D_\mu\phi
-\frac{1}{2}\partial^4 A^\mu\partial_4A_\mu
+D^\mu\phi\,\partial^4A_\mu\Bigr)
\nonumber\\
\label{4.5}
\end{eqnarray}

Expanding the fields above in Fourier harmonics in a similar way to
Sec. II, we get

\end{multicols}
\renewcommand{\theequation}{4.\arabic{equation}}
\Lrule

\begin{equation}
S=\int d^4x\,\mbox{tr}\,\sum_{n=-\infty}^\infty
\Bigl[-\,\frac{1}{4}F_{(n)}^{\mu\nu}F_{(-n)\mu\nu}
-\frac{1}{2}\Bigl(\frac{2n\pi}{R}A_{(n)}^\mu
+i\,D^\mu\phi_{(n)}\Bigr)
\Bigl(\frac{2n\pi}{R}A_{(-n)\mu}
-i\,D_\mu\phi_{(-n)}\Bigr)\Bigr]
\label{4.6}
\end{equation}


\noindent
where

\begin{eqnarray}
&&(D^\mu\phi)_{(n)}=\partial^\mu\phi_{(n)}
+i\sum_{m=-\infty}^{+\infty}
[\phi_{(m)},A^\mu_{(n-m)}]
\nonumber\\
&&F_{(n)}^{\mu\nu}=\partial^\mu A^\nu_{(n)}
-\partial^\nu A^\mu_{(n)}
-i\sum_{m=-\infty}^{+\infty}[A^\mu_{(m)},A^\nu_{(n-m)}]
\label{4.7}
\end{eqnarray}

It is interesting to observe that the non-Abelian character of the
action (\ref{4.6}) induces a gauge structure where  the modes mixed
among themselves. In other words, the gauge multiplet and the modes
form a nontrivial structure that has to be considered as a whole to
preserve the symmetry of the action. Actually, the transformations
(\ref{4.3}) have their mode expanded version given by

\begin{eqnarray}
&&\delta A_{(n)}^\mu=(D^\mu\Lambda)_{(n)}
\nonumber\\
&&\delta\phi_{(n)}=\frac{i2n\pi}{R}\Lambda_{(n)}
-i\sum_{m=-\infty}^{+\infty}\left[\Lambda_{(m)},A^\mu_{(n-m)}\right]
\label{4.9}
\end{eqnarray}

\noindent
where the covariant derivative is defined in (\ref{4.7}). As a
consequence of the above equations,


\begin{eqnarray}
&&\delta F^{\mu\nu}_{(n)}=-i\sum_{m=-\infty}^{+\infty}
\Bigl[\Lambda_{(m)},F^{\mu\nu}_{(n-m)}\Bigr]
\nonumber\\
&&\delta\Bigl(\frac{2n\pi}{R}A^\mu_{(n)}
+i(D^\mu\phi)_{(n)}\Bigr)
=-i\sum_{m=-\infty}^{+\infty}
\Bigl[\Lambda_{(m)},\frac{2(n-m)\pi}{R}
A^\mu_{(n-m)}+i(D^\mu\phi)_{(n-m)}]
\label{4.10}
\end{eqnarray}

\bigskip
\Rrule
\begin{multicols}{2}

\noindent
which is a symmetry of action (\ref{4.6}), as can be verified. It is
interesting to observe that the role played by the St\"uckelberg
fields here is different from the usual one presented by the
non-Abelian Proca model \cite{Kunimasa}. There, the St\"uckelberg
field is introduced by hand in order to just give
$\delta[m\,A^\mu+i(D^\mu\phi)]=0$.

\section{Conclusion}

In this paper we have considered the spontaneous compactifications of
one and two-form gauge theories
from a $D=5$ with a compact dimension to the usual $D=4$
Minkowski spacetime. We have focused our attention to the gauge
invariance of the theories formulated at $D=5$, which should be kept
along the compactification procedure. As usual, there arise massive
modes  in the process of compactification. In principle, the gauge
invariance of these modes could be lost. We observe, however, that
generalized St\"uckelberg fields naturally emerge in order to keep the
content of the original gauge invariance. Although in the Abelian
cases the St\"uckelberg fields correspond to those already found in
the literature, the compactification of the Yang-Mills theory reveals
a new structure of compensating fields. Because of the nonlinearity of
the action, the gauge structure displayed by the mode expansion of
covariant derivatives, curvature tensors and gauge transformations
play a remarkable feature in mixing Fourier modes and gauge multiplet
components in a nontrivial way.

\vspace{1cm}
\noindent
{\bf Acknowledgment:} This work is supported in part by Conselho
Nacional de Desenvolvimento Cient\'{\i}fico e Tecnol\'ogico
- CNPq (Brazilian Research agency) with the support of PRONEX
66.2002/1998-9.

\end{multicols}

\vspace{1cm}
\begin{thebibliography}{30}
\bibitem[a]{}e-mail: {\tt amorim@ if.ufrj.br}
\bibitem[b]{}e-mail: {\tt barcelos@ if.ufrj.br}
\bibitem[]{}
\bibitem{Kaluza} T. Kaluza, Akad. Wiss. Phys. Math. K1 (1921) 966; O.
Klein, Z. Phys. 37 (1926) 895.
\bibitem{Castellani} For a more general reference on compactification
mechanism, see for example, L. Castellani, R. D'Auria and P. Fr\'e,
{\it Supergravity and Superstrings -- A Geometric Perspective} (World
Scientific, 1991), and references therein.
\bibitem{St} E.C.G. St\"uckelberg, Helv. Phys. Acta 30 (1957) 209.
\bibitem{Lahiri} A. Larihi, Phys. Rev. D55 (1997) 5045; D.S. Hwang and
C.-Y Lee, J. Math. Phys. 38 (1997) 30; J. Barcelos-Neto, A. Cabo and
M.B.D. Silva, Z. Phys. C72 (1996) 345.
\bibitem{Dienes} See also K.R. Dienes, E. Dudas, and T. Gherghetta,
Nucl. Phys. B537 (1999) 47, Appendix C.
\bibitem{BFFT} I.A. Batalin and E.S. Fradkin, Phys. Lett. B180 (1986)
157; Nucl. Phys.  B279 (1987) 514; I.A. Batalin, E.S. Fradkin, and
T.E. Fradkina, {\it ibid.} B314 (1989) 158; B323 (1989) 734; I.A.
Batalin and I.V. Tyutin, Int. J. Mod. Phys. A6 (1991) 3255.
\bibitem{Kalb} M. Kalb and P. Ramond, Phys. Rev. D9 (1974) 2273.
\bibitem{Henneaux} For a review see M. Henneaux, Phys. Rep. C126
(1985) 1. A more complete perspective can be found in M. Henneaux and
C. Teitelboim, {\it Quantization of gauge systems} (Princeton Univ.
Press, 1992).
\bibitem{Barc1} J. Barcelos-Neto, J. Math. Phys. 41 (2000) 6661.
\bibitem{Bizdadea} C. Bizdadea and S.O. Saliu, Phys. Lett. B368 (1996)
202.
\bibitem{Kunimasa} T. Kunimasa and T. Goto, Prog. Theor. Phys. 37
(1967) 452.
\end{thebibliography}
\end{document}